\documentclass[12pt]{article}

\def\beq{\begin{equation}}   \def\eeq{\end{equation}}
\def\bea{\begin{eqnarray}}  \def\eea{\end{eqnarray}} \def\nn{\nonumber}
 
\def\lsim{\raise0.3ex\hbox{$<$\kern-0.75em\raise-1.1ex\hbox{$\sim$}}}
\def\gsim{\raise0.3ex\hbox{$>$\kern-0.75em\raise-1.1ex\hbox{$\sim$}}}
\usepackage{epsfig}
\usepackage{wrapfig}

\begin{document}
\setlength{\unitlength}{1mm}

\begin{titlepage}

\begin{flushright}
LPT-Orsay 03-99\\
IPPP/03/76\\
DCPT/03/152\\
\end{flushright}
\vspace{1.cm}

\begin{center}
\vbox to 1 truecm {}
{\large \bf Isolated photon + jet photoproduction as a tool to 
constrain the gluon distribution in the proton and the photon}
\par \vskip 3 truemm
\vskip 1 truecm {\bf M. Fontannaz$^{(a)}$, G. Heinrich$^{(b),}$\footnote{Address 
after December 1, 2003: II. Institut f\"ur Theoretische Physik,
Universit\"at Hamburg,
Luruper Chaussee 149,
22761 Hamburg, Germany.}} 
\vskip 3 truemm

{\it $^{(a)}$ Laboratoire de Physique Th\'eorique, UMR 8627 CNRS,\\
Universit\'e
Paris XI, B\^atiment 210, 91405 Orsay Cedex, France}\\

\vskip 3 truemm

{\it $^{(b)}$ IPPP, Department of Physics, University of Durham, \\Durham DH1 
3LE, England}

\vskip 2 truecm

\normalsize

\begin{abstract}
We analyse how the reaction $\gamma \,p \to \gamma$ + jet + X
can serve to constrain the gluon distributions. 
Our results are based on a code of partonic event generator type
which includes full NLO corrections. 
We conclude that there are phase space domains in
which either the gluon in the photon or the gluon in the proton give
important contributions to the cross section, which should be
observable in HERA experiments.
\end{abstract}

\vspace{3cm}

\end{center}

\end{titlepage}

\section{Introduction}\label{intro}

Over the past years, the 
ZEUS~\cite{Breitweg:1999su,Chekanov:2001aq} 
and H1~\cite{h1} collaborations 
at HERA have been able to observe the photoproduction of
large-$p_{T}$ photons, and the comparisons of data with existing NLO QCD 
predictions~\cite{Gordon:1997yt,Fontannaz:2001ek,Fontannaz:2001nq,Krawczyk:2001tz,Zembrzuski:2003nu} 
appear successful. In photoproduction
reactions, a quasi-real photon, emitted at small angle from the
electron, interacts with a parton from the proton. The photon can
either participate directly in the hard scattering or be resolved
into a partonic system, in which case the parton stemming from the
photon takes part in the hard interaction. Therefore photoproduction is
a privileged reaction to measure or constrain the parton distributions
in the photon and in the proton. In this paper we shall investigate the
possibility to constrain both the gluon in the photon and the gluon in
the proton by looking at the production of a large-$p_{T}$ photon
and a jet.
With the aim of enhancing the contribution of processes involving
initial gluons, we will explore various kinematical domains in detail.
We shall show that there are kinematical configurations which are
dominated either by the gluon in the photon, or by the gluon in the
proton, and which should be accessible to experiment. \par

The photoproduction of large-$p_{T}$ particles and jets has a long
story\,; it offers interesting tests of QCD and gives access to the
measurement of the initial state parton distributions and the final
state fragmentation functions (for a review, see e.g.~\cite{klasen}). 
Reactions involving
large-$p_{T}$ jets and/or hadrons have been copiously observed
because of their large cross sections, whereas the production of prompt
photons has been measured only more recently and the statistical errors are
still rather large. However, this latter reaction has advantages with respect
to jet or hadron production. Indeed a large transverse momentum is
necessary to unambiguously define a jet and to avoid too large 
hadronisation  
and underlying events corrections. The theoretical predictions 
also are subject to uncertainties due to  scale
variations. All these effects are quite sizeable, even for
$E_{T}^{\rm jet} > 21$~GeV \cite{Adloff:2003nr}. 
The photoproduction of
large-$p_{T}$ hadrons also suffers from sizeable theoretical uncertainties  
coming from a large sensitivity to scale variations
and from the fragmentation functions which are not very accurately
measured \cite{Kniehl:2000hk,Fontannaz:2002nu}. 
In the case of a large-$p_{T}$ photon, there
is of course no problem due to jet definition, hadronisation
or inaccurate fragmentation functions (only isolated photons are observed).
Even more importantly, the theoretical predictions are reasonably stable under 
scale variations. Therefore the photoproduction of prompt photons 
appears as an ideal reaction to test QCD and measure the
non-perturbative inputs. \par

However one must keep in mind two reserves, one of experimental 
and another one of theoretical nature. First, the measured cross sections
are small and become rapidly very low at peripheral regions 
of the phase space which might be physically interesting. 
Moreover, the detection of a photon
among the huge amount of large-$p_{T}$ $\pi^0$'s is not an easy task.
Second, only isolated photons are observed, the isolation criterion
being that very little hadronic energy is contained in a cone
surrounding the photon. This has the advantage of reducing the
fragmentation component of the cross section, but at the same time this
isolation can eliminate events containing too much
hadronic energy coming from the underlying event in the cone, 
an effect which cannot be taken into account in the NLO calculations. 
This point has been discussed in
\cite{Fontannaz:2001nq} and studied by the H1 
collaboration \cite{h1,Lemrani:2003mj}.\par

In order to constrain the kinematics of the different subprocesses,
it is important to observe a large-$p_{T}$ jet in association with
the photon, which introduces uncertainties in the comparison between
data and theory due to hadronisation and underlying event phenomena.
However, the effect of these phenomena can be considerably reduced if
the transverse momentum of the jet, which is not well measured, is not
used to constrain the kinematics, but only its rapidity. This is done
by using the variables $x_{LL}$ \cite{Aurenche:2000nc,Fontannaz:2001nq} 
instead of the commonly used $x_{obs}$.
% initially defined by the ZEUS collaboration \cite{}. 
We will make an extensive use of the variables $x_{LL}$ for the 
proton and the photon in order to constrain
the kinematical region relevant for the observation of the gluon
distributions. \par

Our paper is organised in the following way.
In Section 2, we discuss theoretical issues related to the subsequent 
numerical studies, 
such as photon isolation, suitable observables to study the parton 
distributions and the importance of asymmetric cuts on the minimum transverse
energies. Section 3 contains the numerical results, where first the sensitivity 
to the gluon content of the proton is studied. Then we turn to the 
gluon distribution in the real photon before we conclude in Section 4.

\section{Theoretical framework}

As the general framework of the calculation already has been described in
detail in~\cite{Fontannaz:2001ek,Fontannaz:2001nq}, 
we will sketch the method only briefly here and focus instead on 
issues related to the gluon distributions. 

In photoproduction, the electron acts like a source of 
quasi-real photons whose spectrum can be described by the 
Weizs\"acker-Williams approximation, which we use in the following form  
\begin{equation}
f^e_{\gamma}(y) = \frac{\alpha_{em}}{2\pi}\left\{\frac{1+(1-y)^2}{y}\,
\ln{\frac{Q^2_{\rm max}(1-y)}{m_e^2y^2}}-\frac{2(1-y)}{y}\right\}\;.
\label{ww}
\end{equation}
As already mentioned, the quasi-real photon then either takes part 
{\em directly} in the 
hard scattering process, or it acts as a composite object, being a 
source of partons which take part in the hard subprocess. 
The latter mechanism is referred to as {\em resolved} process and 
is parametrised by the parton distributions in the photon  
$F_{a/\gamma}(x^{\gamma},Q^2)$. Thus the distribution of partons 
of type "$a$" in the electron is a convolution
\begin{equation}
F_{a/e}(x_e,M)=\int_0^1 dy \,dx^{\gamma}\,f^e_{\gamma}(y) \,
F_{a/\gamma}(x^{\gamma},M)\,\delta(x^{\gamma}y-x_e)
\label{resolved}
\end{equation}
where in the "direct" case the parton $a$ is the photon itself, i.e. 
$F_{a/\gamma}(x^{\gamma},M)=\delta_{a\gamma}\delta(1-x^{\gamma})$. 
The parton distributions in the photon $F_{a/\gamma}(x,Q^2)$ 
behave like $\alpha/\alpha_s(Q^2)$ for large $Q^2$. Therefore
the additional power of $\alpha_s$ contained in the "resolved" 
component as compared to the "direct" one is compensated. 
This means that the NLO corrections to the resolved component 
can be numerically sizeable and have to be taken into account. 

The cross section can symbolically be written as 
\begin{eqnarray}
&&d\sigma^{e p \to \gamma \, j}(P_e,P_p,P_{\gamma},P_{j})=\nn\\
&&\sum_{a,b}\int dx_e\int d x_p\,
F_{a/e}(x_e,{M})F_{b/p}(x_p,{M})\{d\hat\sigma^{\rm{dir}}\;+ 
\;d\hat\sigma^{\rm{frag}}\}\label{sigma}\\
%\quad + {\small {\cal O}(1/Q^{\nu})}\\
&&\nn\\
&&d\hat\sigma^{\rm{dir}}=d\hat\sigma^{ab\to \gamma\,j}
(x_a,x_b,P_{\gamma},P_{j},\mu,M,M_F)\nn\\
&&d\hat\sigma^{\rm{frag}}=\sum_c\int dz\,D_{\gamma/c}(z,{M_F})
d\hat\sigma^{ab\to c\,j}
(x_a,x_b,P_{\gamma}/z,P_{j},\mu,{M},{M_F})\nn
\end{eqnarray}
where we have split the hard scattering cross sections $\hat\sigma$
explicitly into a "direct" and a "fragmentation" part in order to 
point out that there are two contributions to the "prompt photon" 
in the final state: The "direct" one, where the final state photon 
is produced directly in the hard interaction, and the one where the 
photon stems from the fragmentation of a large-$p_{T}$ 
quark or gluon in the final state. 
This fragmentation process is described by
the fragmentation functions $D_{\gamma/c}(z,{M_F})$. 
At next-to-leading order, 
the distinction between "direct" and  "fragmentation" becomes 
scheme dependent because the final state collinear singularity 
appearing in a  
"direct" process like $\gamma g  \rightarrow \gamma q\bar q$
when a quark becomes collinear to the photon is 
absorbed at the fragmentation scale $M_F$ into the "bare" fragmentation
functions, and where to attribute the finite parts is a matter of 
choice of the factorisation scheme. In our calculation we use the 
${\overline{\rm MS}}$ scheme. 

Note that the cross section (\ref{sigma}) depends on three scales, 
the renormalisation scale $\mu$, the initial state factorisation scale $M$, 
and the fragmentation scale $M_F$, and that it can be considered as 
consisting of 4 categories of subprocesses, depending on whether there is a 
"direct" photon in the initial and/or final state: 1. direct direct, 
2. resolved direct, 3. direct fragmentation, 4. resolved fragmentation. 
Each of these contributions consists of several partonic subprocesses 
at NLO and is strongly scale dependent. Only in the sum this 
scale dependence cancels to a large extent, and only the sum 
can be considered as a physical quantity. This has to be kept in mind 
when the contributions of certain subprocesses only will be considered 
below, in order to estimate the contribution of gluon initiated processes. 
The study of particular subprocesses can be very useful to 
get an idea of the underlying parton dynamics, but cannot be considered 
as precise quantitative statements because of the scale and scheme 
dependence outlined above. 

We would like to emphasize that we calculated the full NLO 
corrections to all four categories of subprocesses. 
We also included the quark loop box contribution $\gamma g\to \gamma g$ 
which is NNLO from a naive $\alpha_s$ power counting point of view, 
but as the process $\gamma g\to \gamma g$ does not exist 
at tree level, the box is the "leading order" diagram 
for this process, and its numerical contribution is quite 
sizeable~\cite{Fontannaz:2001ek}. 
%In \cite{Krawczyk:2001tz,Zembrzuski:2003nu}, the asymptotic behaviour 
%$\sim 1/\alpha_s(Q^2)$ of the photon parton distributions  
%and of the photon fragmentation functions is not considered as 
%a compensation of the factor $\alpha_s$ present in the LO 
%resolved respectively fragmentation contributions. 
The calculation presented in \cite{Krawczyk:2001tz,Zembrzuski:2003nu} 
has no higher order corrections to the resolved-direct, direct-fragmentation
and resolved-fragmentation contributions. It 
only contains the higher order corrections to the
direct-direct part, and the box contribution is also included.

\subsection*{Photon Isolation}
In prompt photon measurements, the experimental challenge consists in
the separation of prompt photon events from the large background of 
secondary photons produced by the decay of light mesons, predominantly 
$\pi^0$ mesons. The latter, when produced at high energy, decay into two 
almost collinear photons which cannot be resolved in the calorimeter. 
However, they are in general accompanied by hadronic energy
and thus this background can be  suppressed by isolation cuts. 

Commonly a cone isolation criterion is used, defined in the following 
way\footnote{A more sophisticated criterion  has been 
proposed in \cite{frixione}, in which the veto on accompanying
hadronic transverse energy is the more severe, the closer the corresponding
hadron is to the photon direction. It has been designed to make the
fragmentation contribution vanish completely, in an infrared safe way,
but is less straightforward to implement experimentally.}:
A photon is isolated if, inside a cone centered around the photon direction 
in the rapidity and azimuthal angle plane, the amount of hadronic transverse 
energy
$E_T^{had}$ deposited is smaller than some value $E_{T,\rm{max}}$ fixed by the
experiment:
\begin{equation}\label{criterion}
%\left.
\begin{array}{rcc} 
\left(  \eta - \eta^{\gamma} \right)^{2} +  \left(  \phi - \phi^{\gamma} \right)^{2}  
& \leq  & R_{\mathrm{exp}}^{2} \\
E_T^{had} & \leq & E_{T,\rm{max}}\;.
\end{array}
%\right\} 
\end{equation}
Following the HERA conventions, we used 
$E_{T,\rm{max}}=\epsilon \,p_T^{\gamma}$ with 
$\epsilon=0.1$ and $R_{\mathrm{exp}}$ = 1.
Of course isolation not only reduces the background from 
secondary photons, but also 
substantially  reduces the fragmentation components, such that the total cross 
section depends very little on the fragmentation functions. 

But another, undesired, effect of isolation is a partial suppression of
the direct contribution. Indeed, the hadronic transverse energy 
$E_T^{had}$ deposited in the cone may stem from the soft underlying event 
due to the spectator-spectator collisions. This renders the effective 
isolation cut more stringent and leads to a decrease of the cross section, 
included the direct contributions.  The simulation of this effect, discussed 
in \cite{Fontannaz:2001ek}, requires a good knowledge of the hadron
distributions in the underlying event, which is asymmetric in $\gamma\,p$ 
collisions\,; it has been studied in recent H1 
publications \cite{h1,Lemrani:2003mj}.

\subsection{Suitable observables to study the parton distributions}

As  observables which serve to reconstruct the longitudinal 
momentum fraction of the parton stemming from the  
proton respectively photon, it is common to use  
\beq
x_{obs}^{\rm{p}}=\frac{p_T^\gamma\,e^{ \eta^\gamma}+
E_T^{\rm jet}\,e^{\eta^{\rm jet}}}{2E^{\rm{p}}}\;,\;
x_{obs}^{\gamma}=\frac{p_T^\gamma\,e^{- \eta^\gamma}+
E_T^{\rm jet}\,e^{-\eta^{\rm jet}}}{2E^{\gamma}} \;.
\label{xobs}
\eeq
However, as the measurement of $E_T^{\rm jet}$ can be a 
substantial source of systematic errors at low $E_T$ values, 
we propose a slightly different variable which does not depend on 
$E_T^{\rm jet}$, 
\beq
x_{LL}^{\rm{p},\gamma}=\frac{p_T^\gamma\,(e^{\pm \eta^\gamma}+e^{\pm
\eta^{\rm jet}})}{2E^{\rm{p},\gamma}}\;.
\label{xll}
\eeq
At leading order, for the non-fragmentation contribution, the variables 
$x_{obs}$ and $x_{LL}$ coincide, and they are also equal to the 
"true" partonic longitudinal momentum fraction, i.e. the argument 
of the parton distribution function. 
At NLO, the real corrections involve 3 partons in the final state (with
transverse momenta $p_{T3}, p_{T4}, p_{T5}$), one of
which -- say parton 5 -- is unobserved. Therefore, $x_{obs}$ and $x_{LL}$
will be different at NLO. 

%\begin{figure}[htb]
\begin{wrapfigure}[20]{r}[0.cm]{10.7cm}
\vspace*{-1.3cm}
%\begin{center}
\epsfig{file=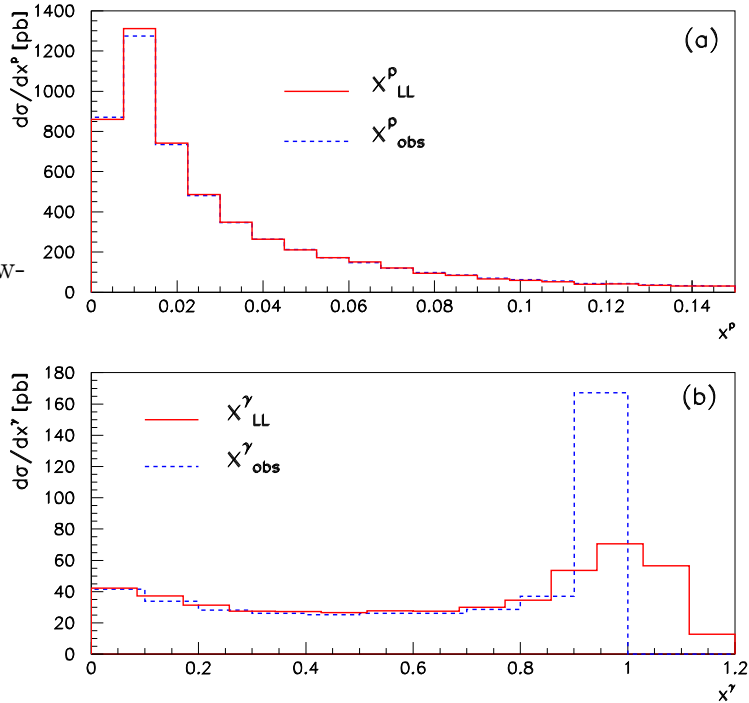,height=10.7cm}
%\end{center}
\caption{Comparison of $x_{LL}$ and $x_{obs}$ for the proton and the photon.
The photon and jet rapidities and transverse energies have been integrated 
in the range 
$-2\leq \eta^{\gamma,{\rm jet}}\leq 4$,  $p_{T}^{\gamma}\geq 6$\,GeV,  
$E_{T}^{\rm jet}\geq 5$\,GeV, $\sqrt s=318$\,GeV.}
\label{xobsll}
\end{wrapfigure}
The difference between $x_{LL}$ and $x_{obs}$ is 
rather small in the proton case, as shown in Fig.~\ref{xobsll}\,a). 
In the photon case, $x_{LL}^\gamma$ and $x_{obs}^\gamma$ 
are very similar in the region $0\,\lsim \,x^\gamma\,\lsim \,0.85$. 
However, for $x^\gamma$ close to one, there are important differences 
between $x_{LL}^\gamma$ and $x_{obs}^\gamma$, the former leading 
to a smoother distribution $d\sigma/dx^\gamma$ if the size of the bins around 
$x^\gamma=1$ is not chosen too small 
(see Fig.~\ref{xobsll}\,b). 

\clearpage

\subsection{Asymmetric cuts}

\begin{wrapfigure}[20]{r}[0.cm]{9cm}
\vspace*{-1.cm}
%\begin{figure}[htb]
%\begin{center}
\mbox{\epsfig{file=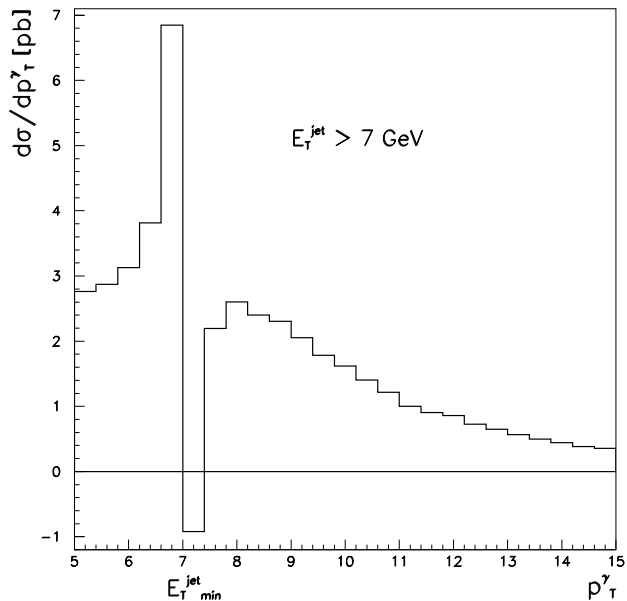,height=9cm}}
%\end{center}
\caption{Logarithmic singularity in $d\sigma/dp_{T}^{\gamma}$ at 
$p_{T}^{\gamma}=E^{\rm jet}_{T,\rm{min}}$. }
\label{etmin}
\end{wrapfigure}

\begin{wrapfigure}[19]{r}[0.cm]{8.cm}
\vspace*{-0.8cm}
\epsfig{file=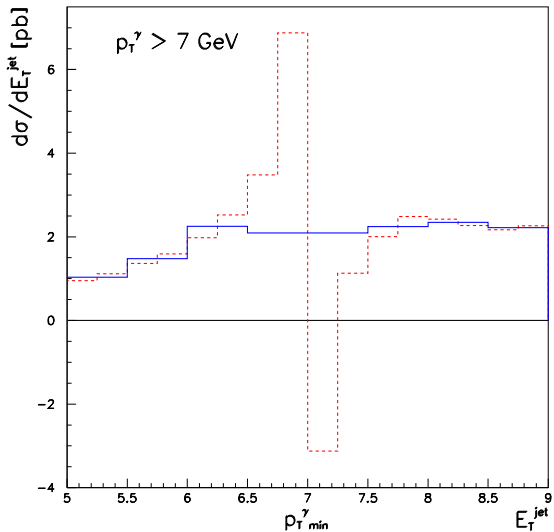,height=8cm}
\caption{The logarithmic singularity in $d\sigma/dE_{T}^{\rm jet}$ at 
$E^{\rm jet}_{T}=p_{T,\rm{min}}^{\gamma}$ is averaged over by choosing 
the binning large enough.}
\label{largebin}
\end{wrapfigure}

It is well 
known\,\cite{Frixione:1997ks, Aurenche:2000nc, Fontannaz:2001nq} that  
symmetric cuts on the minimum transverse energies of dijets or a photon plus 
a jet should be avoided as they amount to including
a region where the fixed order 
perturbative calculation shows infrared sensitivity. 
As explained in detail in~\cite{Fontannaz:2001nq}, the problem stems 
from terms $\sim \log^2(|1-p_T^{\gamma}/E^{\rm jet}_{T,\rm{min}}|)$
which become large as $p_T^{\gamma}$ approaches $E^{\rm jet}_{T,\rm{min}}$, 
the lower cut on the jet transverse energy.
Therefore the partonic NLO cross section has a singular behaviour at 
$p_{T}^{\gamma}=E^{\rm jet}_{T,\rm{min}}$, which is displayed in 
Fig.~\ref{etmin}. Of course, analogously, there are  
$ \log^2(|1-E_T^{\rm jet}/p^{\gamma}_{T,\rm{min}}|)$ terms which become 
large for $E_T^{\rm{jet}}\to p^{\gamma}_{T,\rm{min}}$, see Fig.~\ref{largebin}. 
%the same is true if we fix 
%$E^{\gamma}_{T,\rm{min}}$ and consider the limit 
%$E_T^{\rm{jet}}\to E^{\gamma}_{T,\rm{min}}$. 

The comparison with data can be done in two ways. 
First, if one wants to display e.g. the differential cross section  
$d\sigma/dE_T^{\rm{jet}}$ while 
$p_{T,\rm{min}}^{\gamma}$ lies within the considered $E_T^{\rm{jet}}$ range, 
the binning in 
$E_T^{\rm{jet}}$ must be chosen large enough to average over the logarithmic
singularity which is integrable. 
For instance, the binning 
$E_T^{\rm{jet}}-\Delta\leq p_{T,\rm{min}}^{\gamma}\leq E_T^{\rm{jet}}+\Delta$
with $\Delta=0.5$\,GeV for the bin around $p_{T,\rm{min}}^{\gamma}=7$\,GeV
in Fig.~\ref{largebin} should lead to a correct 
average of the theoretical singularity and allow for a comparison with data. 
Obviously, $d\sigma/dp_T^{\gamma}$ will not exhibit a problem as long as
$E_{T,\rm{min}}^{\rm{jet}}<p_{T,\rm{min}}^{\gamma}$ since the critical point 
$p_{T}^{\gamma}=E_{T,\rm{min}}^{\rm{jet}}$ will not be reached in this case.

Second, one often would like to have a more inclusive cross section such as 
$d\sigma/d\eta^\gamma$, obtained by integrating the differential cross 
section over  $p_T^{\gamma}$ and  $E_T^{\rm{jet}}$. 
In this case one should not choose 
$p_{T,\rm{min}}^{\gamma}=E^{\rm jet}_{T,\rm{min}}$ as this amounts to 
integrating the spectrum of Fig.~\ref{largebin} to the right-hand side of 
$p_{T,\rm{min}}^{\gamma}$ only and thus to picking up only the singular negative
contribution to the NLO cross section, without the compensation coming 
from the positive contribution to the left of $p_{T,\rm{min}}^{\gamma}$. 
As a result, the theoretical prediction, although being finite, 
is infrared sensitive as a consequence of choosing symmetric cuts.
This point has been discussed in detail in ref.~\cite{Fontannaz:2001nq}. 
\clearpage

\begin{wrapfigure}[20]{l}[0.cm]{10.2cm}
\vspace*{-0.5cm}
\epsfig{file=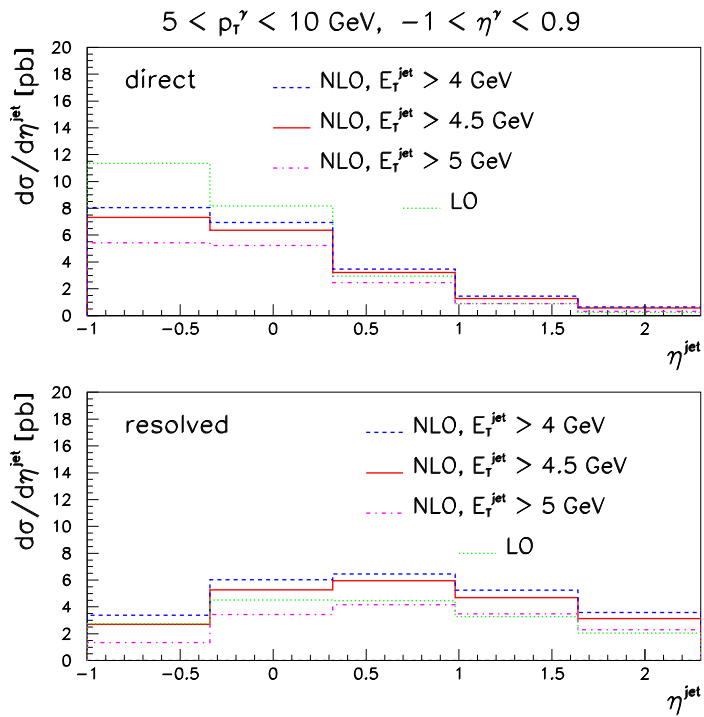,height=10.2cm}
\caption{Magnitude of higher order corrections to direct/resolved parts separately}
\label{etajet}
\end{wrapfigure}

It is also illustrated in Fig.~\ref{etajet},  
where we consider the cross section $d\sigma/d\eta^{\rm jet}$ in a kinematic 
range studied by H1. 
To exhibit the effect of different cuts on $E^{\rm jet}_{T}$, 
we vary $E^{\rm jet}_{T,\rm{min}}$ to take the values 4, 4.5 or 5 GeV, 
while $p_{T,\rm{min}}^{\gamma}$ has been fixed to 5 GeV, and display 
direct and resolved parts separately. 
%\footnote{
Note that the leading order prediction is the same
for all three values of $E^{\rm jet}_{T,\rm{min}}$ as $E^{\rm jet}_{T}$ 
cannot become smaller than $p_{T,\rm{min}}^{\gamma}$ at leading order. 
One observes that in the
direct part at small rapidities, the higher order corrections are
large and negative. 
In the case of symmetric cuts, they are even negative in the resolved part 
at small rapidities, and the theoretical prediction depends strongly 
on the small change in  $E^{\rm jet}_{T,\rm{min}}$ from 
5 GeV to 4.5 GeV, whereas away from the symmetric cut region, the 
prediction is much more stable under small changes of  
$E^{\rm jet}_{T,\rm{min}}$. 
%This illustrates once again the importance of avoiding symmetric cuts

In summary, we repeat that for a successful comparison of data and NLO theory, 
one has to either ensure to stay away from IR singular domains or to consider 
suitably averaged observables. 
In what concerns the cuts on $p_{T}^{\gamma}$ and $E_{T}^{\rm jet}$, 
this amounts to choosing asymmetric cuts adapted to the observable. 
Therefore we disagree with the statement made in~\cite{Chyla:2003in} that 
asymmetric cuts are not superior to symmetric ones.

\clearpage

\section{Numerical results}

Our studies are based on the program 
{\tt EPHOX}\footnote{The program together with detailed documentation 
is available at\\ 
http://wwwlapp.in2p3.fr/lapth/PHOX\_FAMILY/main.html.}, which is a 
partonic Monte Carlo event generator. 
Unless stated otherwise, we use the following input for our numerical 
results: 
A center of mass energy $\sqrt{s}=318$\,GeV with $E_e=27.5$\,GeV and 
$E_p=920$\,GeV is used. The cuts on the minimum transverse energies
of photon and jet are $E_T^{\rm jet}>5$\,GeV, $p_T^{\gamma}>6$\,GeV. 
The rapidities have been 
integrated over in the domain $-2\leq \eta^\gamma,\eta^{\rm jet}\leq 4$ 
unless stated otherwise. 
For the parton distributions in the
proton we take the MRST01~\cite{mrst01} parametrisation, for the photon we 
use AFG04\footnote{This parametrisation has been used in 
ref.~\cite{Fontannaz:2002nu} under the name AFG02.}~\cite{Aurenche:1994in} 
distribution functions and BFG~\cite{Bourhis:1998yu} fragmentation functions. 
We take $n_f=4$ flavours, and for $\alpha_s(\mu)$ we use an exact 
solution of the two-loop renormalisation group
equation, and not an expansion in log$(\mu/\Lambda)$. 
The default scale choice is $M=M_F=\mu=p_T^{\gamma}$. 
Jets are defined using the $k_T$-algorithm~\cite{ktalgo}. 
The rapidities refer to the $e\,p$ 
laboratory frame, with the HERA
convention that the proton is moving towards positive rapidity.

\subsection{The gluon distribution in the proton}

As explained already in Section \ref{intro}, an accurate knowledge of
the gluon distribution in the proton is very important 
at the LHC because of its large gluon luminosity. 
At present the error on important cross sections at the LHC stemming 
from the gluon pdfs is about 5-7\%, but can be up to 20\% 
in certain cases. 
Therefore our aim is to find a region where 
1) the sensitivity to the gluon in the proton is enhanced, 
2) $x^{\rm p}$ is rather large, 
3) the uncertainty stemming from the poorly known gluon in the {\it photon} 
is minimised. 

\subsection*{Cut on $x^\gamma$}
\begin{figure}[htb]
\begin{center}
\epsfig{file=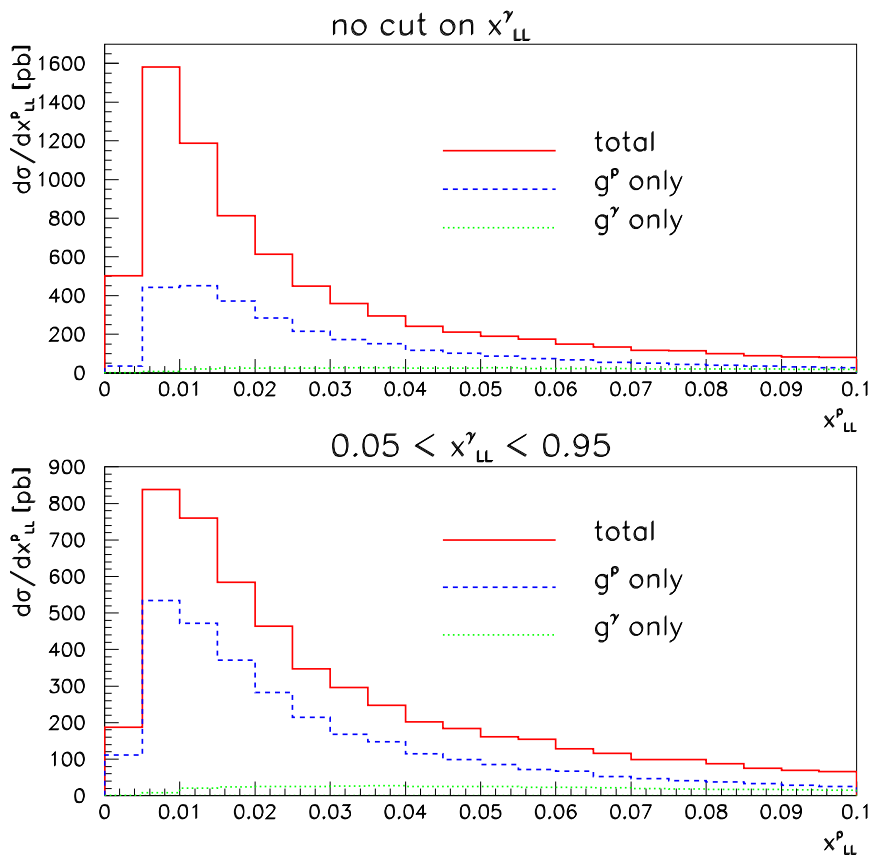,height=12.5cm}
\end{center}
\caption{Effect of a cut on $x^\gamma_{LL}$ to enhance the contribution of 
$g^p$ initiated subprocesses }
\label{figcutxga}
\end{figure}

Requirement 3) can be assured most easily by imposing a lower cut 
$x^\gamma_{min}$ on $x^\gamma$ 
because at small $x^\gamma$ the gluon in the photon is large. 
On the other hand, large values of $x^\gamma$, corresponding 
mainly to direct initial state photons, correspond to small 
values of  $x^{\rm p}$ at a fixed $p_T$ value, according to eq.~(\ref{xll}). 
Therefore 2) can be achieved by a cut $x^\gamma_{max}$ on $x^\gamma$.
In prompt photon production, the contribution from the process 
$\gamma\,(direct) + g^p \to \gamma\,(direct)\,+$\,jet is rather small 
anyway because this process does exist only at NLO. 
Therefore the subprocess $q^\gamma+g^p \to \gamma\,+$\,jet is the one 
which should dominate in the region fulfilling the requirements 1)\,--\,3),   
and our aim is to enhance the contribution of this subprocess. 
To this aim we investigated how cuts on $x^\gamma$ act in this respect, 
and found that the cut $0.05< x_{LL}^{\gamma} <0.95$ 
maximally enhances the sensitivity to the gluon in the proton while 
keeping the contribution from $g^\gamma$ negligible, 
as shown in  Fig.~\ref{figcutxga}. 
Note that the contributions `$g^p$ only' and `$g^\gamma$ only' are not 
disjunct, they both contain the subprocess $g^p+g^\gamma\to \gamma+$\,jet.
This subprocess does not exist at leading order, but beyond LO its 
contribution is non-zero and in fact negative.
This can clearly be seen in Fig.~\ref{figcutxga}: 
The cut $0.05< x^{\gamma}_{LL} <0.95$ actually 
enhances the total value of the $g^p$-initiated part of the cross
section, because the lower limit $0.05< x^{\gamma}_{LL}$ removes mainly
the $g^p+g^\gamma$ initiated part. 
%to a large extent. 
The fact that the $g^p+g^\gamma$ contribution 
is negative also reminds us that the 
individual subprocesses are unphysical, such that these considerations 
can be viewed only as qualitative reflections of the 
underlying parton dynamics. 

\begin{figure}[htb]
\begin{center}
\mbox{\epsfig{file=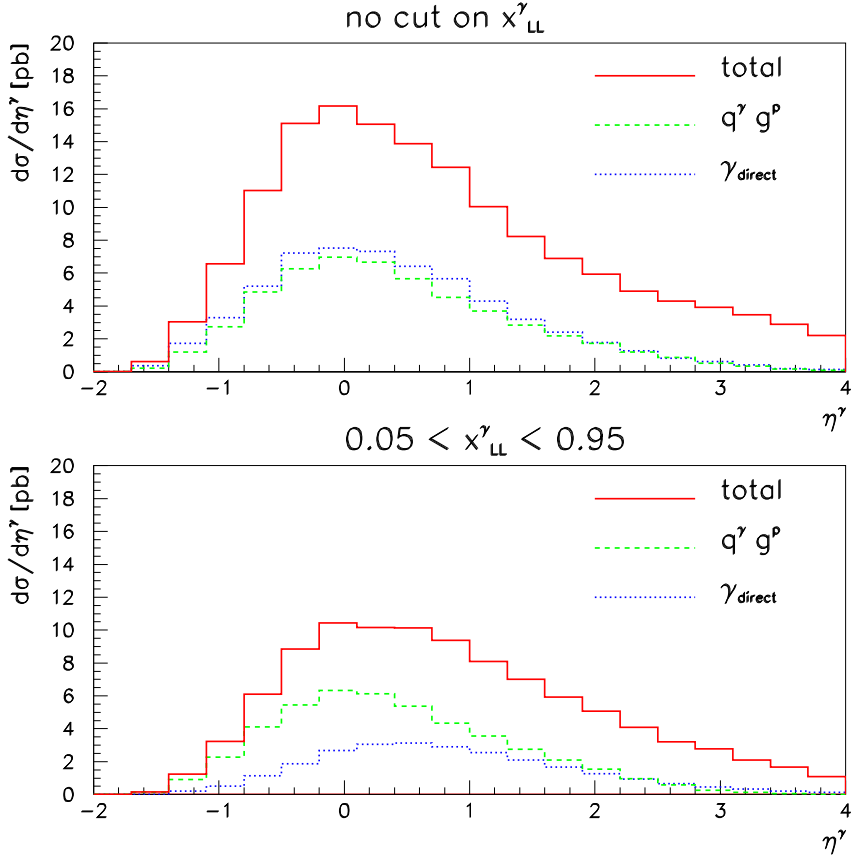,height=12.5cm}}
\end{center}
\caption{ Relative importance of different subprocesses as a function of
photon rapi\-dity. The contribution from the direct photon gets suppressed
by the cut $x^{\gamma}_{LL} <0.95$\,. }
\label{figsubeta}
\end{figure}

Fig.~\ref{figsubeta} shows in detail, as a function of the photon rapidity,   
how the requirement 
$0.05< x^{\gamma}_{LL} <0.95$ suppresses the direct photon contribution 
and enhances the relative importance of the subprocess 
$g^p+q^\gamma\to \gamma+$\,jet, especially in the region 
$\eta^{\gamma}\,\lsim \,1$. 
%The  $\gamma_{direct}g^p$ contribution being negative in the 
%backward region, we again see an {\it enhancement} after the cut. 

\subsection*{Rapidity cuts}
\begin{figure}[htb]
\begin{center}
\mbox{\epsfig{file=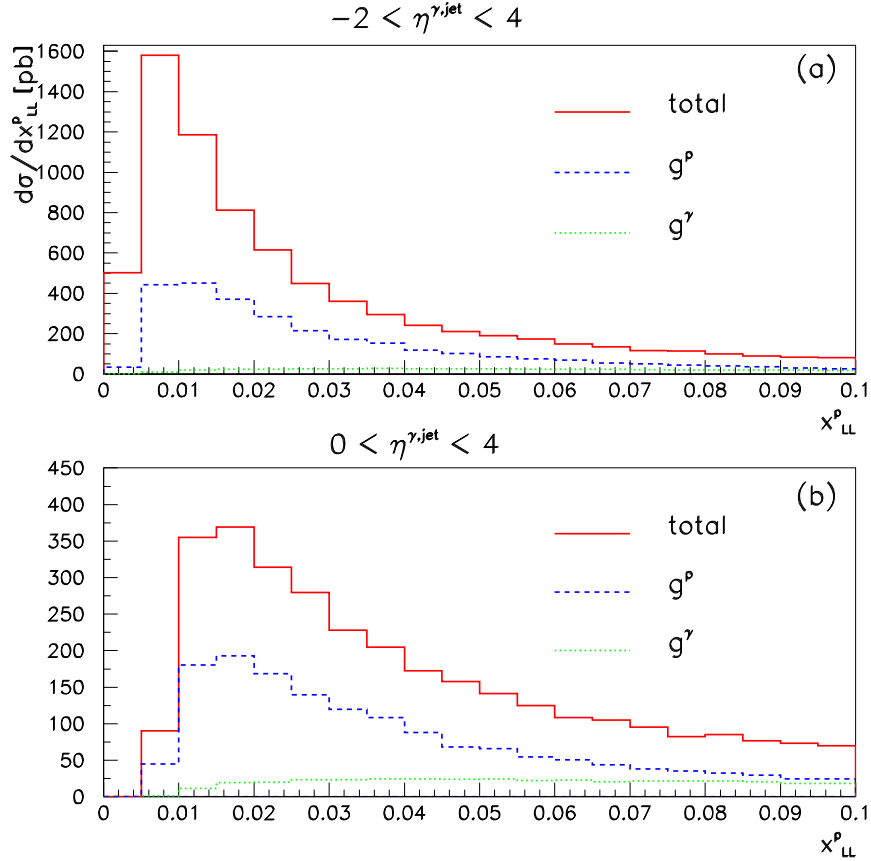,height=12.5cm}}
\end{center}
\caption{Restricting the rapidities to positive values enhances the relative 
contribution of $g^p$-initiated processes.}
\label{xp_rapcuts}
\end{figure}

Another possibility to enhance the contribution of the 
$q^\gamma g^p$ initiated subprocesses is to impose rapidity  cuts. 
Whereas  $x_{obs,LL}$ are variables where rapidity and energy measurements
enter, using 
rapidity cuts only is very straightforward experimentally. 
At small rapidities 
the Compton process $\gamma \,q^p\to \gamma\, q$ dominates. 
Further to the forward region, the importance of 
$q^\gamma g^p$ initiated subprocesses increases, while at 
very large rapidities the gluon in the photon also plays a role. 
Therefore, one can also meet requirements 1) to 3) by restricting the 
photon and jet rapidities to positive values. Fig.~\ref{xp_rapcuts} 
shows that the relative contribution 
of $g^p$-initiated processes  increases from about 35\%
of the total in the full rapidity range 
$-2<\eta^{\gamma,\rm{jet}}<4$ (Fig.~\ref{xp_rapcuts}\,a),   
to about 48\% in the region $0<\eta^{\gamma,\rm{jet}}<4$, 
while the contribution from the gluon 
in the photon is still small, as can be seen from Fig.~\ref{xp_rapcuts}\,b. 
Cutting further (e.g.  $1<\eta^{\gamma,\rm{jet}}<4$) only reduces 
the cross section substantially and introduces a larger 
uncertainty from the gluon in the photon. 

Comparing the two methods, we find that the cut $0.05< x^{\gamma}_{LL} <0.95$ 
reduces the total cross section only by about 31\%, 
enhancing the contribution from the gluon in the proton 
from about 35\% of the total to 56\% of the total.
The rapidity cut $0<\eta^{\gamma,\rm{jet}}<4$ reduces the cross section by 
about 70\% as compared to the range $-2<\eta^{\gamma,\rm{jet}}<4$.

%\clearpage

\subsection*{Scale dependence}

\begin{figure}[htb]
\begin{center}
%\vspace*{-0.8cm}
\epsfig{file=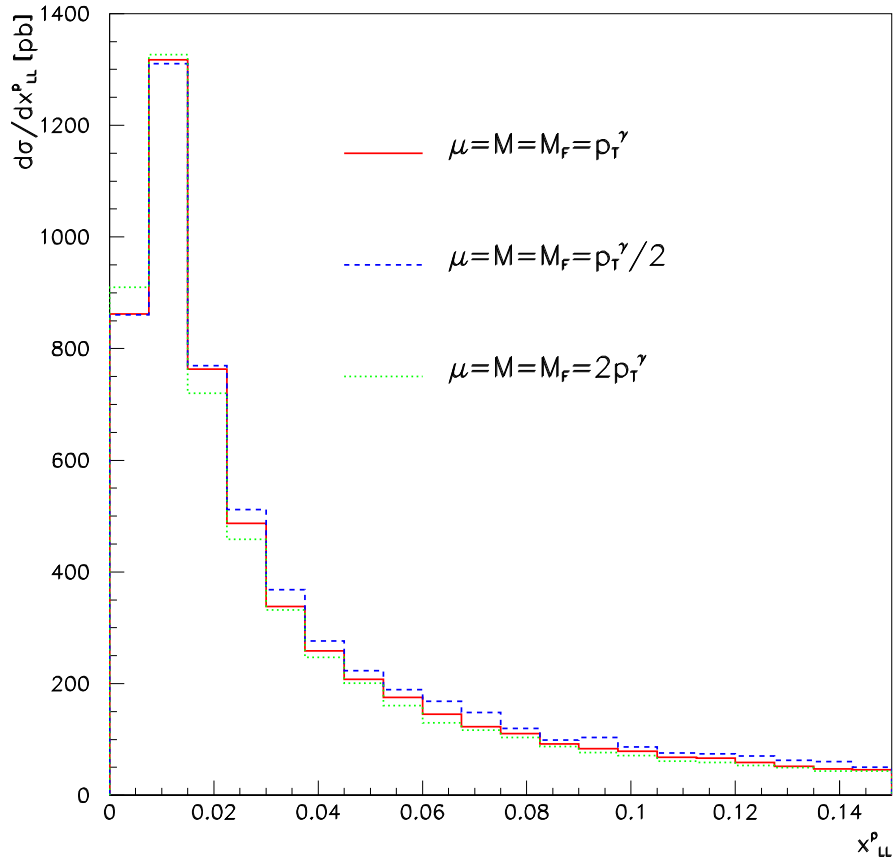,height=13cm}
\caption{Scale dependence of the cross section $d\sigma/dx^p_{LL}$.}
\label{xp_scal}
\end{center}
\end{figure}

\begin{figure}[htb]
\begin{center}
\epsfig{file=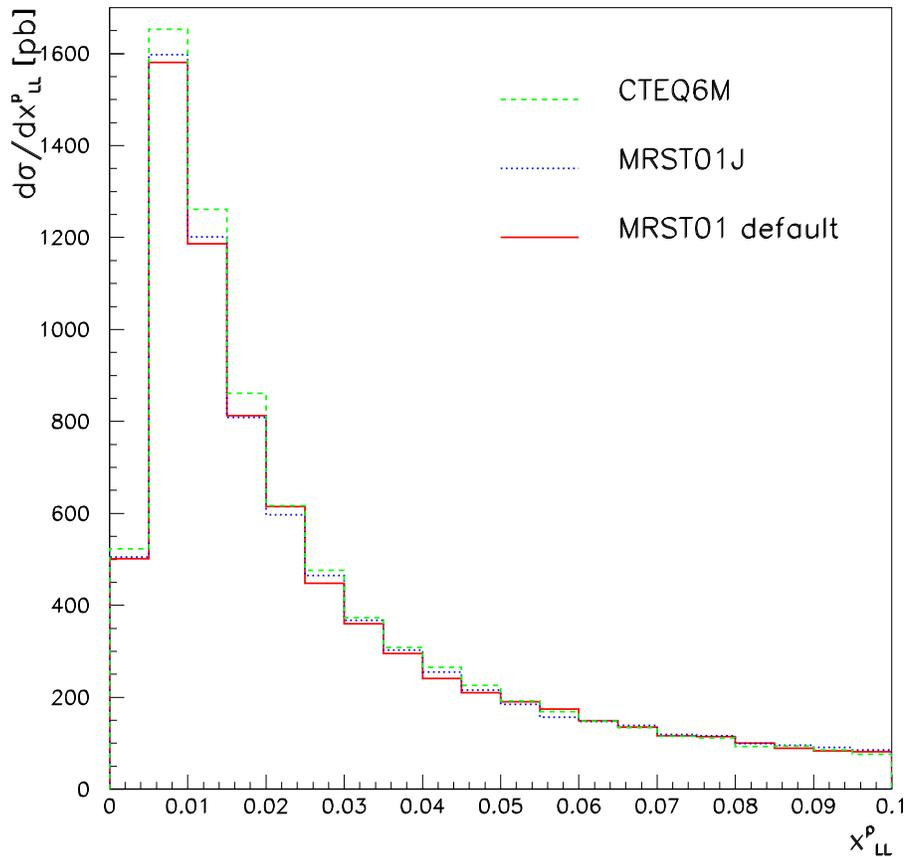,height=13cm}
\caption{Dependence of the cross section $d\sigma/dx^p_{LL}$ on different 
parton distribution functions for the proton.}
\label{mrs4}
\end{center}
\end{figure}

Fig.~\ref{xp_scal} shows that the  NLO cross section 
$d\sigma/dx^p_{LL}$ is very stable under scale changes. 
%\begin{wrapfigure}[22]{r}[0.cm]{10cm}
%\end{wrapfigure} 
\newpage
In Fig.~\ref{mrs4}, we  
show the predictions obtained with different 
parametri\-sations of parton distribution functions for the proton, 
where we chose the set CTEQ6M~\cite{cteq6}
and two different sets of MRST01~\cite{mrst01}, the default set and the set 
MRST01J, which gives better agreement  
with the Tevatron high-$E_T$ inclusive jet data due to a "bump" in the gluon    
distribution at large $x$. 

%\begin{wrapfigure}[20]{r}[0.cm]{13cm}
\begin{figure}[b]
\begin{center}
\epsfig{file=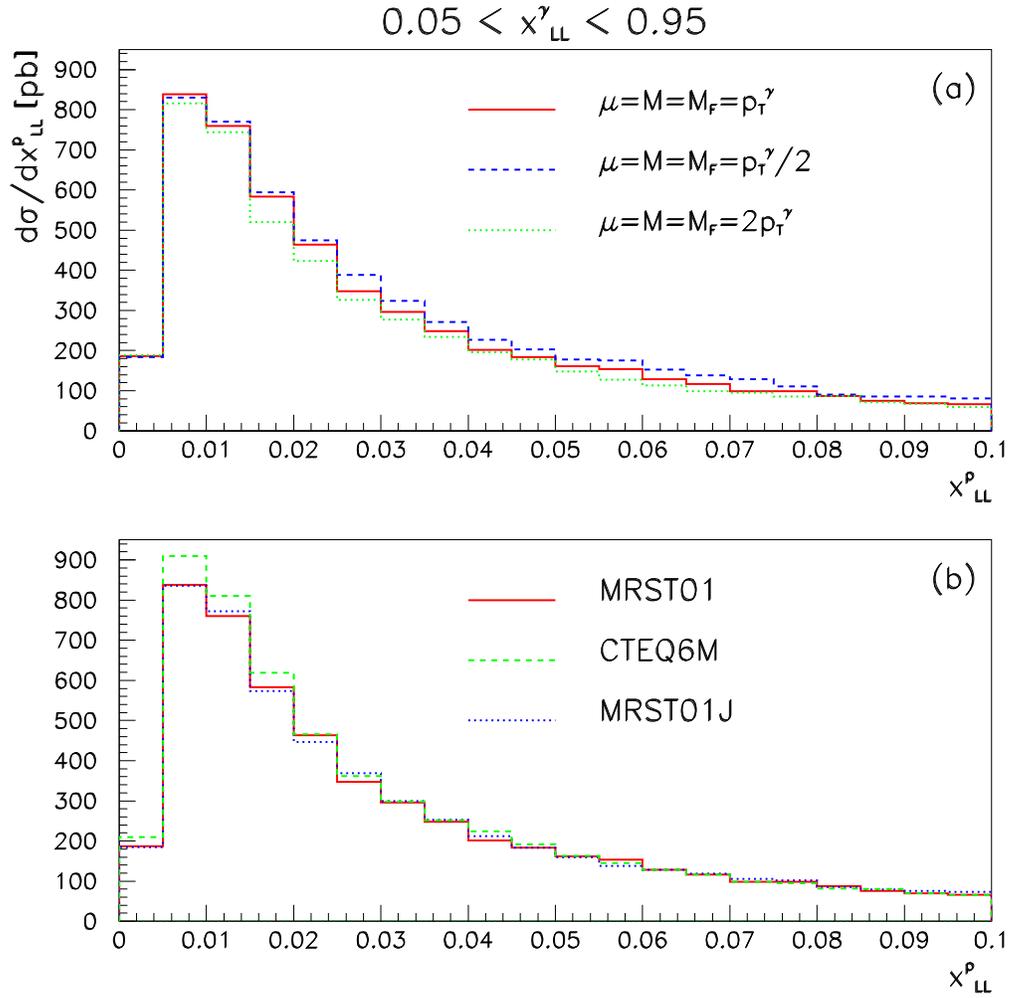,height=14.5cm}
\caption{The cut $0.05< x_{LL}^{\gamma} <0.95$ enhances the gluon contribution 
and thus the differences between the parton distribution functions 
in the region $x_{LL}^{\rm p}\,\lsim\, 0.015$, while it does not 
affect the stability with respect to scale changes.}
\label{pdfscut}
\end{center}
\end{figure}

%\end{wrapfigure} 
Comparing Figs.~\ref{xp_scal} and \ref{mrs4}, we notice 
that the differences in the cross sections due to different 
parton distribution functions are of the
order of the variations due to the scale changes. 
However, this situation can be somewhat improved: 
Fig.~\ref{pdfscut} shows that the cut $0.05< x_{LL}^{\gamma} <0.95$
makes the differences between various 
parametrisations more pronounced  
(as it enhances the gluon initiated 
contribution to the cross section), 
especially in the region 
$x_{LL}^{\rm p}\,\lsim\, 0.02$. 
On the other hand, the variation  
of the cross section due to scale changes
in the region $x_{LL}^{\rm p}\,\lsim\, 0.015$  
%of the order of the numerical error, so  
is not increased  
by the presence of the cut on  $x_{LL}^{\gamma}$.  
Therefore, the reaction 
$\gamma \,{\rm p}\to \gamma$\,+\,jet\,+\,X could indeed 
be useful to further constrain the gluon in the proton
in the range $x_{LL}^{\rm p}\,\lsim\, 0.015$. 
However, it is also clear  that 
data with very high statistics are needed 
to distinguish between different parametrisations. 
\clearpage

\subsection{The gluon content of the photon}

The photoproduction of large-$p_T$ jets, hadrons and photons 
are privileged reactions to explore the gluon content of the 
resolved photon, which is hardly observable in $\gamma\,\gamma^*$ DIS. 
However, as discussed in the introduction, the scale dependence of the hadron
and jet production cross sections is not negligible, whereas the photon cross
section is more stable. This fact should allow us a more accurate determination
of the gluon distribution in the photon, $g^\gamma(x^{\gamma},Q^2)$. 
%\begin{wrapfigure}[23]{l}[0.cm]{10cm}
%\vspace*{-0.5cm}
\begin{figure}[htb]
\begin{center}
\epsfig{file=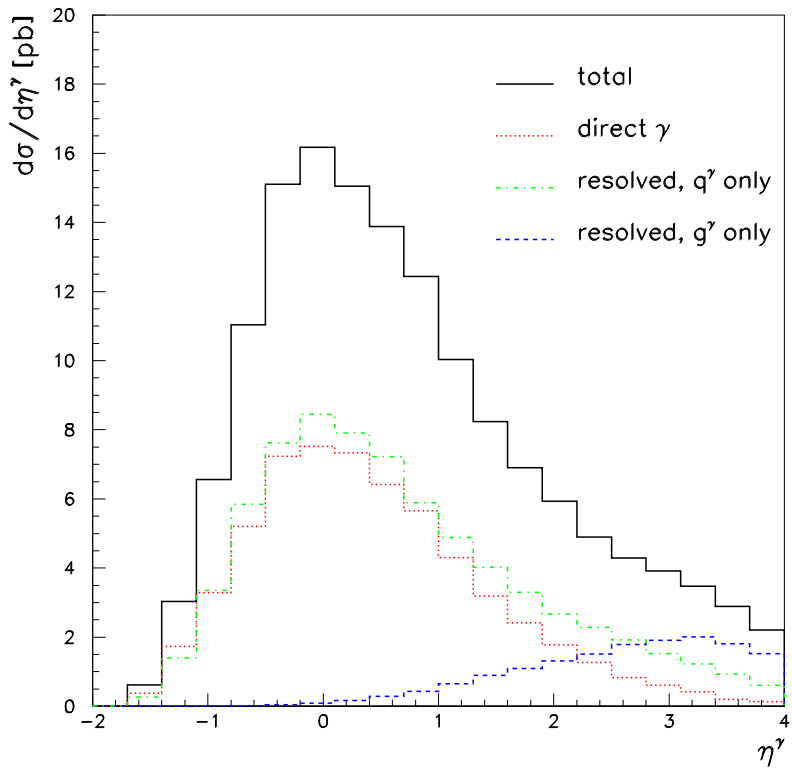,height=11.5cm}
\caption{Magnitude of different subprocesses over the full photon rapidity
range. The jet rapidities have been integrated over 
$-2<\eta^{\rm jet}<4$, and  $E_T^{\rm jet}>5$\,GeV, 
$p_T^{\gamma}>6$\,GeV.}
\label{fulleta}
\end{center}
\end{figure} 

In Fig.~\ref{fulleta} we display the various contributions to the cross section
$d\sigma/d\eta^\gamma$. The gluon distribution  $g^\gamma(x^{\gamma},Q^2)$ only contributes at
small values of $x^{\gamma}$, corresponding to large values of $\eta^\gamma$, and we 
shall try, by various cuts, to enhance the re\-la\-tive contribution of this
component.

\begin{figure}[htb]
\begin{center}
\epsfig{file=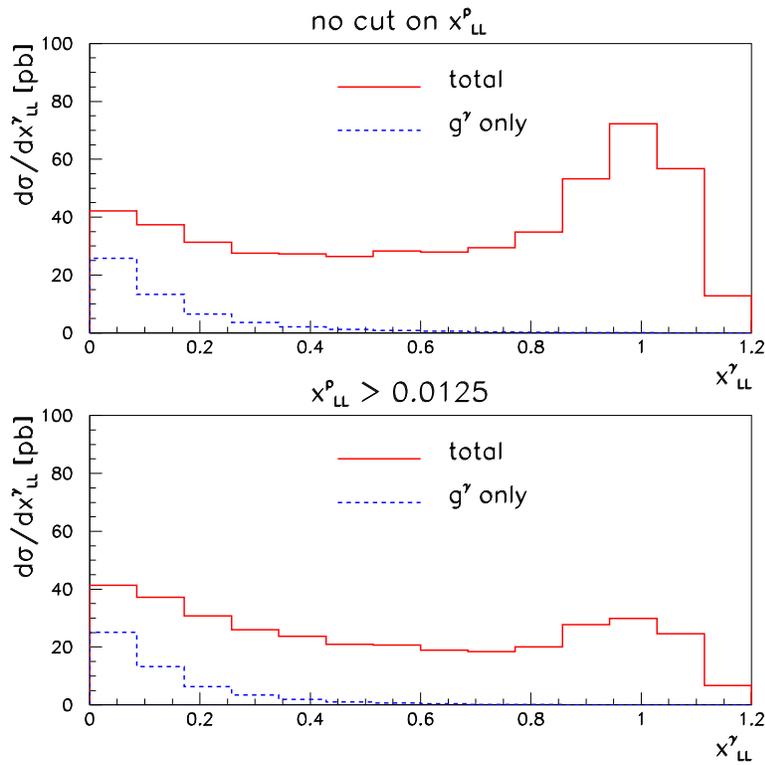,height=11cm}
\caption{Effect of a lower cut on $x_{LL}^{\rm p}$ on the relative contribution 
from the gluon in the photon.}
\label{cutxp}
\end{center}
\end{figure}

\begin{figure}[htb]
\begin{center}
\mbox{\epsfig{file=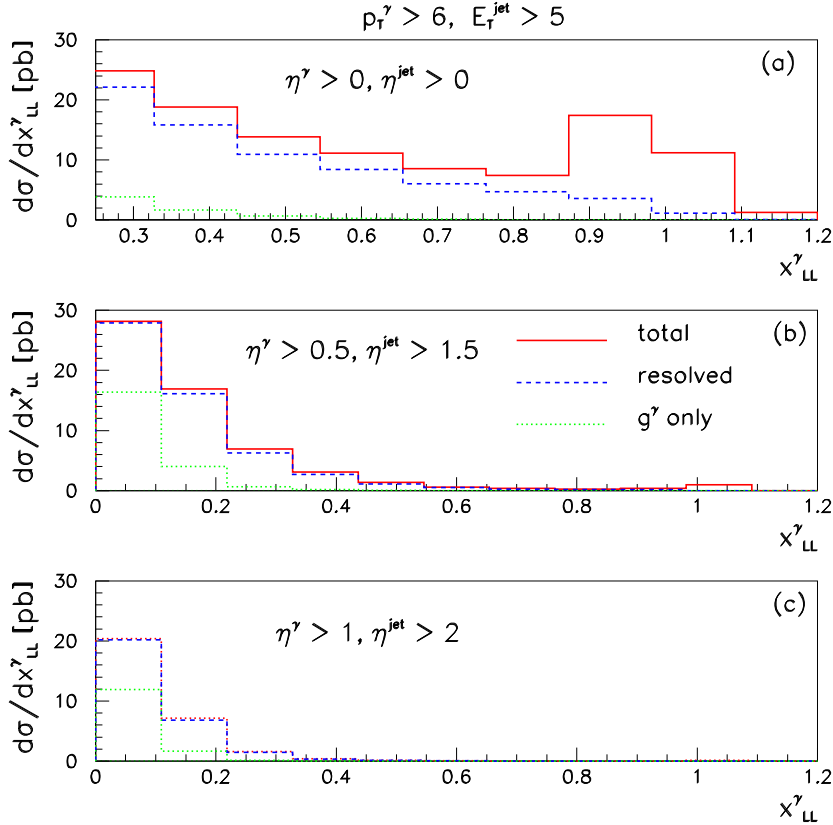,height=12cm}}
\end{center}
\caption{Effect of rapidity cuts to enhance the contribution 
from the gluon in the photon.}
\label{cutrap}
\end{figure}

In Fig.~\ref{cutxp} we see that the direct contribution, 
corresponding to $x_{LL}^{\gamma}$ close to one, does not screen 
the contribution initiated by the gluon in the photon. 
But at smaller values of $x_{LL}^{\gamma}$, the `background' coming from other 
subprocesses, such as $q^{\gamma}g^p\to q g$, is large. 
Exactly this fact has been exploited to enhance the gluon from the proton 
by restricting $x_{LL}^{\gamma}$ to intermediate values, 
see Fig.~\ref{figcutxga}. 
Now we would like to enhance the gluon from the {\it photon}, and therefore we 
impose a lower cut on $x_{LL}^{\rm p}$ in order to reduce the 
contributions from the gluon in the proton.  
However, this cut has no effect  at very small values of 
$x_{LL}^{\gamma}$, where the gluon in the photon is most visible. 
Therefore, contrary to the situation for the gluon in the  proton 
treated in the previous subsection, 
cuts on the photon and jet rapidities are more effective than cuts on 
$x_{LL}^{\rm p}$ to enhance the gluon in the photon, 
as shown in Fig.~\ref{cutrap}.  
%\begin{wrapfigure}[23]{r}[0.cm]{11.5cm}
%\vspace*{-0.5cm}
%\end{wrapfigure}
Fig.~\ref{cutrap}\,a) shows that if the photon and jet rapidities are
restricted to positive values, the resolved photon component is already 
fairly large, but the gluon content of the latter is still small. 
If we restrict the rapidities more to the forward region -- especially the
jet rapidity, which can be measured at larger angles -- the direct 
photon contribution is almost completely suppressed, and the gluon 
contribution makes up almost 40\% of the total, as shown in 
Fig.~\ref{cutrap}\,b).  Imposing even more severe cuts only decreases 
the cross section further without increasing the gluon content 
substantially, as can be seen from Fig.~\ref{cutrap}\,c). 
Therefore the rapidity cut $\eta^{\gamma}>0.5,\,\eta^{\rm jet}>1.5$
seems to be the optimal compromise between enhancement of the gluon content 
and reduction of the cross section. 

\begin{figure}[htb]
\begin{center}
\mbox{\epsfig{file=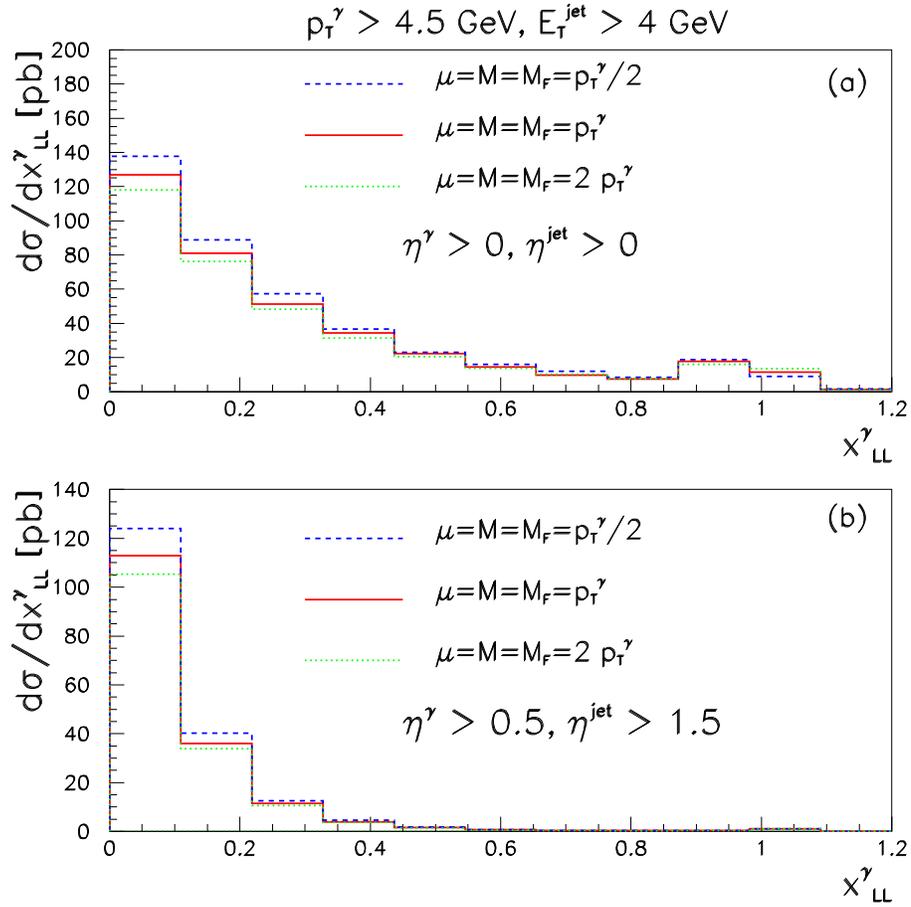,height=13cm}}
\end{center}
\caption{Scale dependence of $d\sigma/dx_{LL}^{\gamma}$ in the presence of
forward rapidity cuts.}
\label{xgamscal}
\end{figure}

Note that the lower cuts on the transverse momenta are rather large,  
$p_T^\gamma>6$\,GeV, $E_T^{\rm jet}>5$\,GeV. 
One can increase the cross section by choosing lower $p_T$ cuts, as 
shown in Fig.~\ref{xgamscal}. 
This figure also shows the scale dependence of 
$d\sigma/dx_{LL}^{\gamma}$ in the presence of the cuts 
$\eta^\gamma>0,\,\eta^{\rm jet}>0$ respectively 
$\eta^\gamma>0.5,\,\eta^{\rm jet}>1.5$. 
The behaviour of the cross section $d\sigma/dx_{LL}^{\gamma}$, 
which varies by $\pm$\,8\,\% under the scale changes, is less good than 
the behaviour of $d\sigma/dx_{LL}^{\rm p}$ (see Fig.~\ref{xp_scal}).
However, one should keep in mind that the distribution $g^\gamma$
is poorly known and that a determination of the latter with an accuracy of 
$\pm$\,10\,\% would already be welcome.

\clearpage

\section{Conclusions}

In this work we studied the possibility to measure the gluon
distribution in the proton and in the
photon by means of the reaction $\gamma$ + p $\to \gamma\, +$ jet + X.
This reaction is well suited for such a study because of 
the stability of the theoretical prediction 
under variations of the renormalisation and factorisation scales, 
and because the prompt photon cross section does not suffer 
from large uncertainties due to hadronisation in the final state. 
\par
                                                                                
We have shown that gluon induced subprocesses give important
contributions to the cross section in the  $x$-ranges
$0 < x_{LL}^{\rm p} \ \lsim \ 0.1$ for the proton and 
$0 < x_{LL}^{\gamma} \ \lsim \  0.2$ for the photon. 

The effects of cuts on $x_{LL}^{\gamma}$ respectively 
$x_{LL}^{\rm p}$, or on the
pseudo-rapidities $\eta^{\gamma}$ and $\eta^{\rm jet}$, 
are investigated in detail. We found that a
judicious choice of cuts allows us to enhance the `signals', i.e. 
the gluon induced subprocesses, 
over the `background' stemming from other subprocesses,    
to constitute up to $\sim 50\,$\% of the total cross section. 
               
However, the relevant cross sections are small, of the
order of 10 - 50 pb. 
Clearly these numbers require a large
luminosity to obtain observable effects.

\vspace*{8mm}       

{\bf\Large Acknowledgements}

\medskip

\noindent
GH would like to thank the LPT Orsay and the LAPTH Annecy for their 
hospita\-li\-ty while part of this work has been completed, 
and also G\"unter Grindhammer for encouragement 
to pick up again the subject of asymmetric cuts.

\end{document}